\documentclass[twocolumn,nofootinbib,aps,superscriptaddress,preprintnumbers]{revtex4}

\usepackage{graphicx}

\input myBabarsym

\newcommand\Abar{\kern 0.18em\overline{\kern -0.18em A}{}}

\renewcommand{\Im}{{\rm Im}}

\newcommand{\nn}{\nonumber}
\newcommand{\beq}{\begin{equation}}
\newcommand{\eeq}{\end{equation}}
\newcommand{\beqa}{\begin{eqnarray}}
\newcommand{\eeqa}{\end{eqnarray}}

\newcommand\deltaAlpha{\Delta\alpha}

\def\Rcpi{\ensuremath{R_{+-}}}
\def\Rnpi{\ensuremath{R_{00}}}

\def\OMIT#1{}

\begin{document}

\preprint{\vbox{\hbox{LAL 05--34} \hbox{LBNL--57484} \hbox{MIT--CTP 3624}
\hbox{hep-ph/0506228} }}

\title{\boldmath Testing the dynamics of $B\to \pi\pi$ and constraints on
$\alpha$}


\author{Yuval Grossman}
\affiliation{Department of Physics, Technion--Israel Institute of Technology,
  Technion City, 32000 Haifa, Israel}
\affiliation{Physics Department, Boston University, Boston, MA 02215}
\affiliation{Jefferson Laboratory of Physics, Harvard University,
  Cambridge, MA 02138}

\author{Andreas H\"ocker}
\affiliation{Laboratoire de l'Acc\'el\'erateur Lin\'eaire,
  IN2P3-CNRS et Universit\'e Paris-Sud, BP 34, F-91898 Orsay Cedex, France}

\author{Zoltan Ligeti}
\affiliation{Ernest Orlando Lawrence Berkeley National Laboratory,
  University of California, Berkeley, CA 94720}

\author{Dan Pirjol$\,$}
\affiliation{Center for Theoretical Physics, MIT, Cambridge, MA 02139}


\begin{abstract}

In charmless nonleptonic $B$ decays to $\pi\pi$ or $\rho\rho$, the ``color
allowed" and ``color suppressed" tree amplitudes can be studied in a systematic
expansion in $\alpha_s(m_b)$ and $\lqcd/m_b$.  At leading order in this
expansion their relative strong phase vanishes.  The implications of this
prediction are obscured by penguin contributions.  We propose to use this
prediction to test the relative importance of the various penguin amplitudes
using experimental data.  The present $B\to\pi\pi$ data suggest that there are
large corrections to the heavy quark limit, which can be due to power
corrections to the tree amplitudes, large up-penguin amplitude, or enhanced weak
annihilation.  Because the penguin contributions are smaller, the heavy quark
limit is more consistent with the $B\to\rho\rho$ data, and its implications may
become important for the extraction of $\alpha$ from this mode in the future.

\end{abstract}

\maketitle

\section{Introduction}

Nonleptonic $B$ decays to light hadrons provide information about $\CP$
violation.  In particular, the decays to $\pi\pi$, $\rho\pi$ and
$\rho\rho$ can determine the weak phase $\alpha$.  The theoretical
challenge is to disentangle the strong interaction physics from the
weak phase one would like to determine. For the decay $\Bz\to\pip\pim$
the \B factories study the \CP asymmetry,
\beqa
\lefteqn{\frac{\Gamma[\Bzb(t)\to \pi^+\pi^-] - \Gamma[B^0(t)\to \pi^+\pi^-]}
              {\Gamma[\Bzb(t)\to \pi^+\pi^-] + \Gamma[B^0(t)\to \pi^+\pi^-]}
        }
         \nn\\[0.2cm]
  &&\hspace{0.5cm}
  = S_{+-} \sin(\Delta m\, t) - C_{+-} \cos(\Delta m\, t)\,,
\eeqa
with the present world averages~\cite{pipiCP,Group:2004cx}
\beq\label{SCdata}
S_{+-} = -0.50 \pm 0.12, \qquad C_{+-} = -0.37 \pm 0.10\,.
\eeq
If the $B \to \pi^+\pi^-$ amplitude were dominated by contributions
with a single weak phase, the observable
\beq\label{aeff}
\sin(2\alpha_{\rm eff}) = S_{+-} \big/ \sqrt{1-C_{+-}^2} \,,
\eeq
would be equal to $\sin2\alpha$ and $C_{+-}$ would be zero. The data indicate
that this is not a good approximation.  An isospin analysis~\cite{Gronau:1990ka}
still allows a theoretically clean determination of $\alpha$ if the
$B^0 \to \pi^0\pi^0$ and $\Bzb \to \pi^0\pi^0$ rates are precisely
measured.  Since this requires very large data samples, several strategies
have been proposed to extract $\alpha$ from $\alpha_{\rm eff}$ relying on
theoretical inputs.

In the last few years the theory of $B \to \pi\pi$ decays has advanced
considerably.  Using the heavy quark limit, factorization theorems have been
proven for the decay amplitudes at leading order in $\Lambda/m_b$.  The
amplitudes in Eq.~(\ref{Apipi}) arise from the matrix element of the effective
Hamiltonian,
\beqa\label{Heff}
&& H_{\rm eff} = - {4G_F\over\sqrt2} \bigg[
  \lambda_u \Big( C_1 O_1^u + C_2 O_2^u
  + \sum_{i\geq3} C_i^c O_i \Big) \nn\\
&&{}\qquad\ + \lambda_c \Big(C_1 O_1^c + C_2 O_2^c
  + \sum_{i\geq3} C_i^c O_i\Big) \nn\\
&&{}\qquad\ + \lambda_t\, \sum_{i\geq3} C_i^t\, O_i \bigg]\,,
\eeqa
where CKM-unitarity was not used, and $i = 3, \ldots, 6, 8$.  (In the usual
notation one has $C_i=C_i^c-C_i^t$.)  Its $\Bb\to \pi\pi$ matrix element can be
parameterized as
\beqa\label{Apipi}
\Abar(\Bzb\to \pi^+\pi^- )
  &=& - \lambda_u (T+P_{u}) - \lambda_c P_c - \lambda_t P_{t} \nn\\
  &=& e^{-i\gamma } T_{\pi\pi} + e^{i\phi} P_{\pi\pi}\,, \nn\\
\sqrt2 \Abar(\Bzb\to \pi^0 \pi^0)
  &=& \lambda_u (-C +P_u) + \lambda_c P_c + \lambda_t P_t \nn\\
  &=& e^{-i\gamma } C_{\pi\pi} - e^{i\phi} P_{\pi\pi}\,, \nn\\
\sqrt2 \Abar(B^-\to \pi^- \pi^0) &=& -\lambda_u(T+C)=e^{-i\gamma } T_{-0} \,,
\eeqa
where $\lambda_q = V_{qb} V_{qd}^*$.  (We neglect isospin
breaking~\cite{isospinbreak} and the contributions of electroweak penguins, the
dominant part of which can be included model independently~\cite{ewp}.)  
In Eq.~(\ref{Apipi}) $T+P_u$ and $C-P_u$ are the $B\to\pi^+\pi^-$ and
$B\to\pi^0\pi^0$ matrix elements of the terms in the first line in
Eq.~(\ref{Heff}), while $P_c$ and $P_t$ are the matrix elements of the second
and third lines, respectively.  This implies that each of the $T+P_u$, $C-P_u$,
$P_c$ and $P_t$ terms are separately renormalization group invariant. 

There is an ambiguity in Eq.~(\ref{Apipi}) related to the freedom in choosing
the weak phase $\phi$, in terms of which the amplitudes are written.  There are
two widely used conventions corresponding to eliminating either $\lambda_t$ or
$\lambda_c$ using unitarity (some aspects of this were discussed in
Refs.~\cite{conv}).  In the t-convention one eliminates $\lambda_t$ from
Eq.~(\ref{Apipi}), while in the c-convention one eliminates $\lambda_c$.
Table~\ref{tab:conv} shows the expressions for the amplitudes and $\phi$ in
these conventions.  Once a choice is made, $T_{\pi\pi}$, $C_{\pi\pi}$,
$P_{\pi\pi}$, and $T_{-0}$ can be extracted from the data, while further
theoretical input is needed to determine $T$, $C$ and $P_{u,c,t}$.  

The amplitudes in Eq.~(\ref{Apipi}) (and their \CP conjugates)
satisfy the isospin relation
\beq\label{ispin}
\frac1{\sqrt2} \Abar(\Bzb\to \pi^+\pi^- )
  + \Abar(\Bzb\to \pi^0 \pi^0) = \Abar(B^-\to \pi^- \pi^0)\,.
\eeq
The ``tree'' amplitudes also satisfy the relation
\beq\label{TT}
T_{\pi\pi} + C_{\pi\pi} = T_{-0}\,,
\eeq
which will play an important role in this paper, and we refer to it as
the ``tree triangle'' (TT).

Expanding the amplitudes in soft-collinear effective theory (SCET)~\cite{scet},
one can define the leading (in $\Lambda/m_b$) parts of $T$, $C$, and $P_u$
separately in terms of matrix elements of distinct SCET operators~\cite{bp},
which we denote with $(0)$ superscripts.  The relative strong phase of $T^{(0)}$
and $C^{(0)}$ is suppressed by $\alpha_s$~\cite{bprs,qcdf}, and therefore
\beq\label{phiTC}
\phi_T \equiv \arg \bigg({T^{(0)} + P_u^{(0)}\over T+C}\bigg)
  = {\cal O}\big[\alpha_s(m_b),\, \lqcd/m_b\big]\,.
\eeq
The numerator includes $P_u^{(0)}$ so that $\phi_T$ is scale independent. The
denominator could be defined to contain $T^{(0)}+C^{(0)}$, and our choice is for
later convenience.  Neither of these affect the right-hand side of
Eq.~(\ref{phiTC}) [recall:  $P_u^{(0)} / T^{(0)} = {\cal O}(\alpha_s)$].  We
define $T^{\prime(0)} \equiv T^{(0)} + P_u^{(0)}$ and $T + P_u \equiv
T^{\prime(0)} + P_u'$, and in the rest of this paper the primes will be
dropped.  Thus, hereafter, $P_u$ contains the power suppressed corrections to
$T+P_u$ (including weak annihilation).

\begin{table}[t]
\caption{The $B\to \pi\pi$ amplitudes and the phase of the penguin
amplitude in the c- and t-conventions ($P_{ij} \equiv P_i-P_j$).}
\label{tab:conv}
\setlength{\tabcolsep}{0.0pc}
\begin{tabular*}{\columnwidth}{@{\extracolsep{\fill}}ccc}
\hline \hline
~~amplitude~~  &  ~~~~t-convention~~~~ & ~~~~c-convention~~~~  \\
\hline
$T_{\pi\pi}$ & $|\lambda_u|(-T - P_{ut})$ & $|\lambda_u|(-T - P_{uc})$ \\
$C_{\pi\pi}$ & $|\lambda_u|(-C + P_{ut})$ & $|\lambda_u|(-C + P_{uc})$ \\
$P_{\pi\pi}$ & $-|\lambda_c|P_{ct}$	&  $|\lambda_t|P_{ct}$ \\ \hline
$\phi$		&  $\pi$ 	&  $\beta$ \\ \hline\hline
\end{tabular*}
\end{table}

The implications of Eq.~(\ref{phiTC}) for the determination of $\alpha$ are
obscured by the fact that $T$ and $C$ are not directly observable. The
amplitudes $T_{\pi\pi}$ and $C_{\pi\pi}$ in Eq.~(\ref{Apipi}) that can be
extracted from the data include contributions from $P_{u,c,t}$.   The heavy
quark limit also determines the power counting for the penguin amplitudes,
however, the convergence of the expansion for the penguins is less clear than it
is for the trees.  At leading order in $\Lambda/m_b$ the calculable parts of
$P_{u,c,t}$ are suppressed by $\alpha_s$ or the small Wilson coefficients
$C_{3,4}$.  At subleading order, the QCD factorization (QCDF) formula for $P_t$
contains sizeable ``chirally enhanced" corrections, comparable to the leading
order term~\cite{qcdf}.  The possible size of nonperturbative contributions to
$P_c$ has also been the subject of debate~\cite{bprs,pcdedbate}.  A large $P_c$
amplitude was found in fits using the leading order factorization results in
SCET~\cite{bprs}, or adding a free parameter to the leading order QCDF
result~\cite{cpengs}.  In QCDF $P_c$ is claimed to be computable at leading
order without nonperturbative inputs, while $P_t$ receives sizable ``chirally
enhanced" ${\cal O}(\Lambda/m_b)$ corrections.  Equation~(\ref{phiTC}) and
allowing for large long distance contribution to $P_c$ was used in
Ref.~\cite{brs} to determine $\alpha$ without using the measurement of $C_{00}$ 
(the direct $\CP$ asymmetry in $B\to \pi^0\pi^0$).

The penguin amplitudes $P_c$ and $P_t$ introduce a difference between the TTs in
the two conventions.  The $P_u$ amplitude is common to $T_{\pi\pi}$ in the t-
and c-conventions, but $P_c$ enters $T_{\pi\pi}$ in the c-convention and $P_t$
enters $T_{\pi\pi}$ in the t-convention.  Understanding the relative hierarchy
of the three penguin amplitudes, $P_{u,c,t}$, is important if one is to use
Eq.~(\ref{phiTC}) for the determination of $\alpha$.  In addition, it may also
shed light on the $\Lambda/m_b$ power counting for the penguin amplitudes.  In
this paper we show that by comparing the shapes of the TT in the c and
t-conventions we can gain empirical knowledge about the relative sizes of $P_u$,
$P_c$ and~$P_t$.

\section{Isospin analysis and tree triangle}

\begin{figure}[t]
\includegraphics[width=0.85\columnwidth]{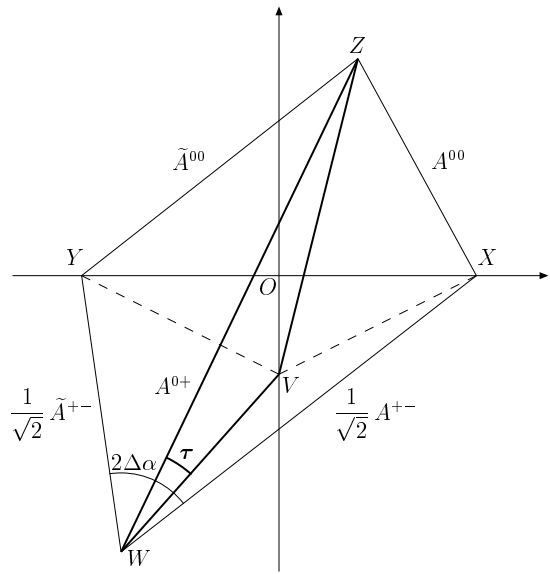}
\caption{Isospin triangles for $\Bb$ and $B$ decay, $WZY$ and $WZX$. $WVZ$
is the tree triangle (TT), Eq.~(\ref{TT}), with $WV = T_{\pi\pi}$ and
$ZV = C_{\pi\pi}$.  The dashed lines show the $P_{\pi\pi}$ amplitudes.}
\label{fig:ispin}
\end{figure}

The isospin relation in Eq.~(\ref{ispin}) holds for both the $\Bb$ and $\B$
decay amplitudes, denoted by $\bar A$ and $A$, respectively. It is convenient to
define $\widetilde A^{ij} = e^{2i\gamma} \Abar^{ij}$, so that $A^{0+} =
\widetilde A^{0-}$.  Figure~\ref{fig:ispin} shows the resulting two isospin
triangles, $WZX$ and $WZY$, where the tree triangle, $WZV$, is also drawn.  We
follow the notation of Ref.~\cite{GLSS}, but normalize $A(B^+\to \pi^0\pi^+) =
\overline{WZ} = 1$.

To determine the TT from the data, recall that the $WZX$ and $WZY$
isospin triangles can be obtained from the direct $\CP$ asymmetries
$C_{+-}$ and $C_{00}$, and the ratios of branching fractions
\beqa\label{Rcn}
\Rcpi &=& \frac{{\cal B}(B^0\to \pi^+\pi^-)}{2\, {\cal B}(B^+\to \pi^+\pi^0)}\,
  \frac{\tau_{B^+}}{\tau_{B^0}} = 0.44 ^{\,+0.07}_{\,-0.06}\,, \nn\\
\Rnpi &=& \frac{{\cal B}(B^0\to \pi^0\pi^0)}{{\cal B}(B^+\to \pi^+\pi^0)}\,
  \frac{\tau_{B^+}}{\tau_{B^0}} = 0.29 ^{\,+0.07}_{\,-0.06}\,,
\eeqa
where we used the experimental inputs from~\cite{pipiBr,Group:2004cx}.  Taking
the ratios eliminates an arbitrary overall normalization parameter. To
determine the coordinates of $V$, however, the measurement of $S_{+-}$
is also needed.

It is convenient to define the coordinates of $X$ and $Y$ to be
$(\pm\ell,0)$, with
\beq\label{ell}
\ell^2 = \frac12 \Rcpi \Big[1 - \sqrt{1-C_{+-}^2} \cos 2\deltaAlpha \Big],
\eeq
where $\deltaAlpha \equiv \alpha-\alpha_{\rm eff}$ and $\alpha_{\rm eff}$ is
defined in Eq.~(\ref{aeff}).  The four coordinates of $W$ and $Z$ and the phase
$\deltaAlpha$ are given by the solutions of the five equations \cite{GLSS}
\beqa\label{WZ}
1 &=& (x_Z - x_W)^2 + (y_Z - y_W)^2, \nn\\
\Rnpi &=& x_Z^2+y_Z^2+\ell^2, \nn \\
\Rcpi &=& x_W^2+y_W^2+\ell^2, \nn \\
\Rcpi C_{+-} &=& -2 \ell x_W,  \nn\\
\Rnpi C_{00} &=& -2 \ell x_Z.
\eeqa
The $XVY$ angle is $2(\phi+\gamma)$, so that the $y$ coordinate of
$V\, (0,y_V)$ is
\beq\label{yV}
y_V = \cases{ - \ell \cot \gamma\,, & \mbox{in the t-convention}\,,\cr
  {\phantom -}\ell \cot \alpha\,, & \mbox{in the c-convention}\,. \cr }
\eeq
Equations~(\ref{WZ}) can be solved for $\deltaAlpha$ and the coordinates of $W$
and $Z$.  Because of the relative orientation of the amplitudes  $A^{+-}$ and
$\widetilde A^{+-}$ adopted in Fig.~1, the solution must also satisfy
$\mbox{sgn}(\deltaAlpha) = \mbox{sgn}(y_W)$.

Some important properties of the solutions are apparent.  First, $x_W=0$ if and
only if $C_{+-}=0$ (similarly, $x_Z=0$ if and only if $C_{00}=0$).  Second, the
sign of $x_W$ ($x_Z$) is opposite of that of $C_{+-}$ ($C_{00}$).  Thus, $WZ$
crosses the $y$ axis if and only if the direct $\CP$ asymmetries in the charged
and neutral modes have opposite signs.

In the rest of this section, we treat the simplified case where $C_{00}$ is not
known.  The first four equations in (\ref{WZ}) can be used to solve for the
coordinates of $W$ and $Z$ as functions of $\deltaAlpha$. For any given value of
$\deltaAlpha$, $W$ and $Z$ are determined up to a two-fold ambiguity,
corresponding to the reflection of $Z$ about the $WO$ line. These equations
also  place bounds on $\ell$ and $\deltaAlpha$~\cite{Grossman:1997jr,GLSS}
\beqa\label{bound}
\ell^2 &\leq& \Rcpi \Rnpi - {(1-\Rcpi-\Rnpi)^2\over4}\,, \nn\\
\cos(2\deltaAlpha) &\geq& {(1+\Rcpi-\Rnpi)^2 - 2\Rcpi \over 2\Rcpi\sqrt{1-C_{+-}^2}} \,.
\eeqa
We refer to these inequalities as the isospin bound, and define $\alpha_{\rm
bound} \equiv \alpha_{\rm eff} \pm \deltaAlpha_{\rm max}$, which can be obtained from
Eqs.~(\ref{aeff}) and (\ref{bound}), and $\gamma_{\rm bound} \equiv
\pi-\beta-\alpha_{\rm bound}$. (Here, and in what follows $ \beta$ is treated as
known.) The coordinates of $W$ and $Z$ at the isospin bound satisfy
\beq
{x_Z \over x_W}\bigg|_{\rm bound} = {y_Z \over y_W}\bigg|_{\rm bound}
  = - {1+\Rnpi-\Rcpi \over 1-\Rnpi+\Rcpi} \,.
\eeq
This means that at the isospin bound $W$, $Z$, and $O$ are on one line
and that at the bound
\beq
C_{00}\big|_{\rm bound} = - {\Rcpi\over \Rnpi}\,
  {1+\Rnpi-\Rcpi\over 1-\Rnpi+\Rcpi}\; C_{+-}\big|_{\rm bound} \,.
\eeq
The present data gives at the isospin bound $C_{00} = -(1.1 \pm 0.1)\, C_{+-}$,
which is almost $2\sigma$ from the measurements of $C_{+-}$ in
Eq.~(\ref{SCdata}) and $C_{00} = -0.28 ^{+0.39}_{-0.40}$~\cite{c00,Group:2004cx}.

In general, and even at the isospin bound, the $V$ vertex of the TT depends on
$S_{+-}$ via Eq.~(\ref{yV}).  Thus, the shape of the TT at the bound is not
fixed, but depends on the experimental results.  This dependence enters through
$\alpha_{\rm eff} + \deltaAlpha$ and implies that if one uses a constraint on
the shape of the TT to extract $\alpha$, then i) the solution is not invariant
under $\deltaAlpha \leftrightarrow -\deltaAlpha$, and ii) the allowed values  of
$\deltaAlpha$ are not the same for each discrete ambiguity of $\alpha_{\rm
eff}$.   Both of these points are different from the well-known symmetry
properties of  the usual isospin analysis.

The theory prediction of a small strong phase in Eq.~(\ref{phiTC}) implies that
the TT should be nearly flat, up to penguin contributions, small $\alpha_s$ and
unknown $\Lambda/m_b$ corrections.  While the penguin contamination makes the
definition of the TT itself convention dependent, it is interesting to consider
under what conditions the TT can be flat, and its relation to the isospin
bound.  Since at the isospin bound $W$, $Z$, and $O$ are on a line, unless
$y_V=0$, the TT is flat at the isospin bound if and only if $x_W=x_Z=0$.  This
implies that if any two of the following statements hold, then the other three
follow:
\beq\label{cute}
\begin{tabular}{l}
1. The t-convention TT is flat for generic $\alpha$; \\
2. The c-convention TT is flat for generic $\alpha$; \\
3. $\alpha$ is at the isospin bound; \\
4. $C_{+-}=0$; \\
5. $C_{00}=0$.
\end{tabular}
\eeq
Equivalently, when one of the statements in (\ref{cute}) holds, the other
four are either all true or all false.  This shows that whether the TT is
flat near the isospin bound or not depends on the value of $\alpha$; {\em i.e.},
the TT being flat and $\alpha$ (or $\gamma$) being close to the isospin
bound are in principle unrelated.

\section{\boldmath Constraints on $\alpha$}

In Ref.~\cite{brs}, the predicted smallness of $\phi_T$ and $P_{ut}$ was used
to imply that the TT in the t-convention is (near) flat, which, in turn, was
used to extract $\gamma$ without the insufficiently known $C_{00}$.
In this section we discuss the implications of knowing an angle in the TT
for the determination of $\alpha$, using a method which makes transparent
the dependence of the constraints on $\alpha$ on the data.

For given $\Rcpi$, $\Rnpi$, and $C_{+-}$, the first four equations in (\ref{WZ})
together with (\ref{ell}) determine the coordinates of $W$ and $Z$ as functions
of $\deltaAlpha$.  If, in addition, an angle in the TT is also known, then the
position of the point $V$ is determined.  We find it simplest to discuss the
constraints in terms of the (convention dependent) observable phase,
\beq\label{taudef}
\tau^{(q)} \equiv \arg \bigg({ T_{\pi\pi}^{(q)} \over T_{-0}}\bigg)
  = \arg \bigg(1 + {P_{uq} \over T^{(0)}}\bigg) + \phi_T ,
\eeq
where $q=c$ or $t$.  The TT is near flat in either convention if $|\tau|\ll 1$.
Note that if the penguin amplitudes vanished, then $\tau^{(t)} = \tau^{(c)} =
\phi_T$.  We can determine the coordinates of $V$ as a function of
$\deltaAlpha$ in two ways: from the value of $\tau$ and the coordinates of $W$
and $Z$
\beq\label{eqdelta-a}
y_V(\deltaAlpha) = y_W - x_W \frac{y_Z-y_W - (x_Z-x_W)\tan\tau}
  {x_Z-x_W + (y_Z-y_W)\tan\tau} \,,
\eeq
and from Eq.~(\ref{yV}) if $\beta$, $S_{+-}$ and $C_{+-}$ are
measured
\beq \label{eqdelta-b}
y_V(\deltaAlpha) = \cases{\ell\, \cot(\beta + \alpha_{\rm eff} + \deltaAlpha)\,,
  & \mbox{t-convention}, \cr
  \ell\, \cot(\alpha_{\rm eff} + \deltaAlpha)\,,
  & \mbox{c-convention}. \cr}
\eeq
The expression in (\ref{eqdelta-b}) is convention dependent, because so is the
definition of $\tau$ that enters in (\ref{eqdelta-a}).  These two equations form
an implicit equation for $\deltaAlpha$.

\begin{figure}[t]
\includegraphics[width=0.9\columnwidth]{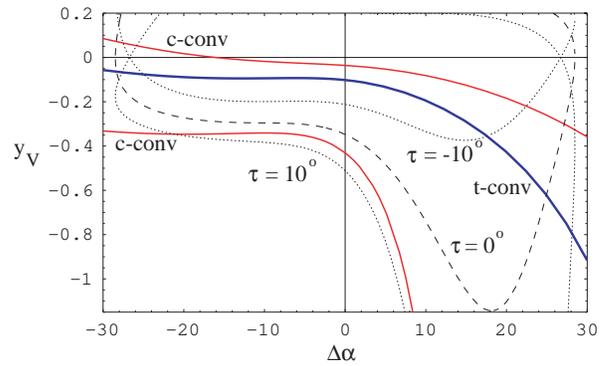}
\caption{The solid curves are $y_V$ vs.\ $\deltaAlpha$ from
Eq.~(\ref{eqdelta-b}): the darker (blue) curve corresponds to the t-convention
and $\alpha_{\rm eff} \simeq 106^\circ$, while the lighter (red) curves to the
c-convention (the upper one for $\alpha_{\rm eff} \simeq 106^\circ$, the lower
one for $\alpha_{\rm eff} \simeq 164^\circ$).  The dashed curve shows the
solution of Eq.~(\ref{eqdelta-a}) for $\tau = 0$, and the dotted curves are
$\tau = +10^\circ$ (lower) and $\tau = -10^\circ$ (upper).}
\label{fig:sol}
\end{figure}

Figure~\ref{fig:sol} illustrates this method for the central values of the
data.  The solid curves show the solution for $y_V(\deltaAlpha)$ vs.\
$\deltaAlpha$ from Eq.~(\ref{eqdelta-b}): the darker (blue) curve corresponds to
the t-convention and $\alpha_{\rm eff} \simeq 106^\circ$, while the lighter
(red) curves correspond to the c-convention (the upper one for $\alpha_{\rm eff}
\simeq 106^\circ$, the lower one for its mirror solution $\alpha_{\rm eff}
\simeq 164^\circ$).  The dashed curve shows $y_V$ vs.\ $\deltaAlpha$ from
Eq.~(\ref{eqdelta-a}) for $\tau = 0$, and its intersections with the solid
curves determine the value of $\deltaAlpha$, which together with $\alpha_{\rm
eff}$ gives $\alpha$.  For the purpose of illustration the dotted curves show
$\tau = +10^\circ$ (lower curve) and $-10^\circ$ (up-most curve).

The $\tau=0$ curve goes to $y_V=0$ at the isospin bound (see
Fig.~\ref{fig:sol}), in accordance with our result in Sec.~II that if
$\deltaAlpha$ is at the isospin bound and the TT is flat, then $y_V=0$.  The
right-hand side of Eq.~(\ref{eqdelta-b}) is small in this region of
$\deltaAlpha$, since the argument of the cotangent is close to $90^\circ$ (the
central values of the $\pi\pi$ data give $\alpha_{\rm eff} \simeq 106^\circ$, so
that at the smallest value of $\deltaAlpha \simeq -28^\circ$, $\beta +
\alpha_{\rm eff} + \deltaAlpha \simeq 102^\circ$ and $\alpha_{\rm eff} +
\deltaAlpha \simeq 79^\circ$).  These two facts imply that there is a solution
for $\deltaAlpha$ near the isospin bound with a flat TT; however, this is a
coincidence and not a necessity.

In Ref.~\cite{brs} it was found that for small $\tau^{(t)}$ the solution for
$\deltaAlpha$ was close to the isospin bound.  This can be easily seen from
Fig.~\ref{fig:sol}.  The dashed and dotted curves are steep near the bound for
negative $\deltaAlpha$, so changing $\tau$ hardly changes the solution for
$\deltaAlpha$.  However, for the other solution (corresponding to positive
$\deltaAlpha$, and a value of $\alpha$ disfavored by the global CKM
fit~\cite{Charles:2004jd}), the error is significantly larger, since the
dependence of $\deltaAlpha$ on $\tau$ is stronger.  The allowed region of
$\deltaAlpha$ is particularly sensitive to $\Rnpi$; for example, for $\Rnpi=0.2$
(which is a bit more than $1\sigma$ lower than its present central value) the
$|\tau| < 10^\circ$ constraint would include almost all values of $\deltaAlpha$
that are allowed by the isospin analysis.  Note that with the current data the
error of $\alpha$ extracted using the  constraint of a small $\tau$ increases
with decreasing $\Rnpi$, contrary to  the isospin analysis.

\begin{figure*}[t]
\includegraphics[width=\columnwidth]{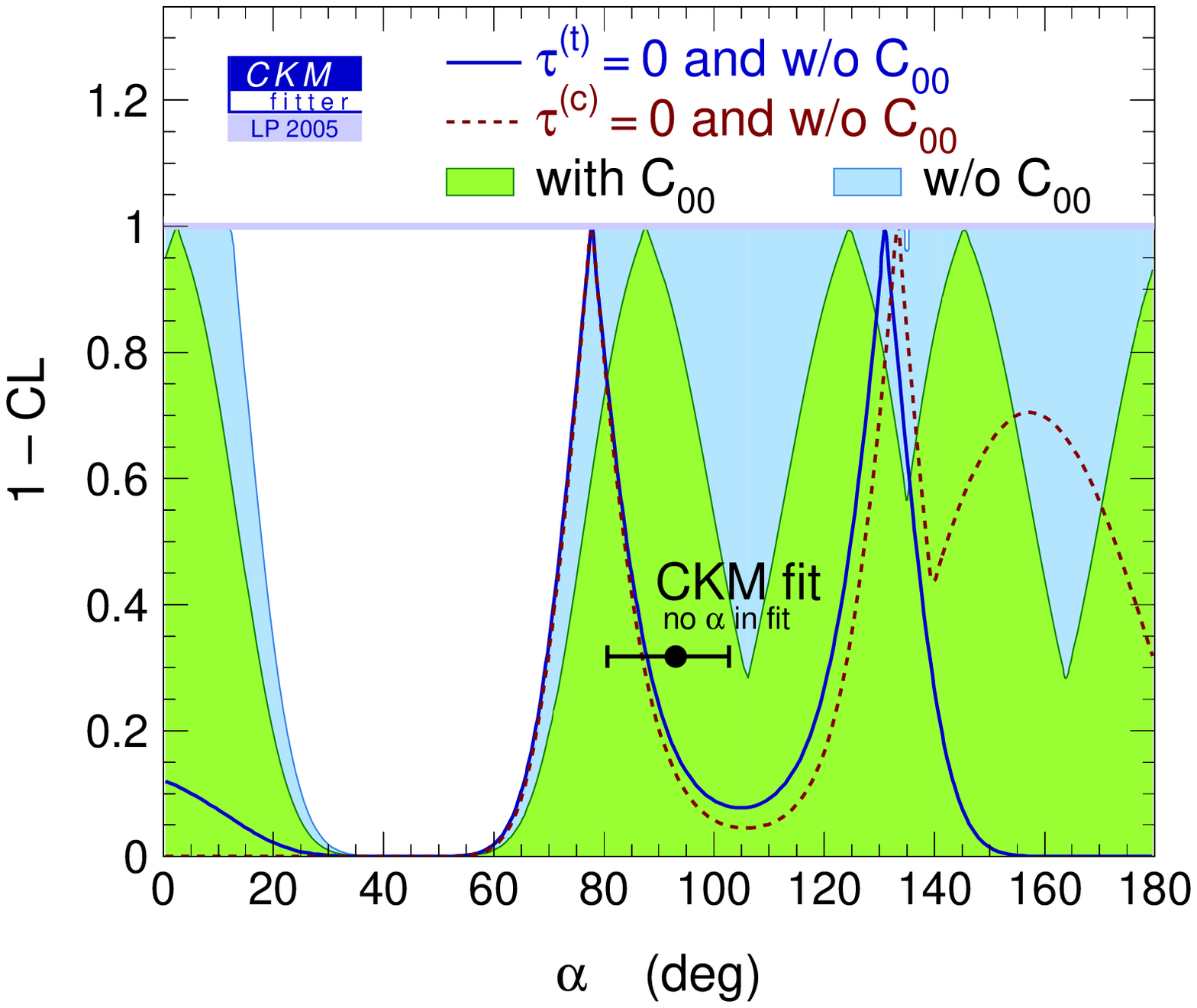} \hfill
\includegraphics[width=\columnwidth]{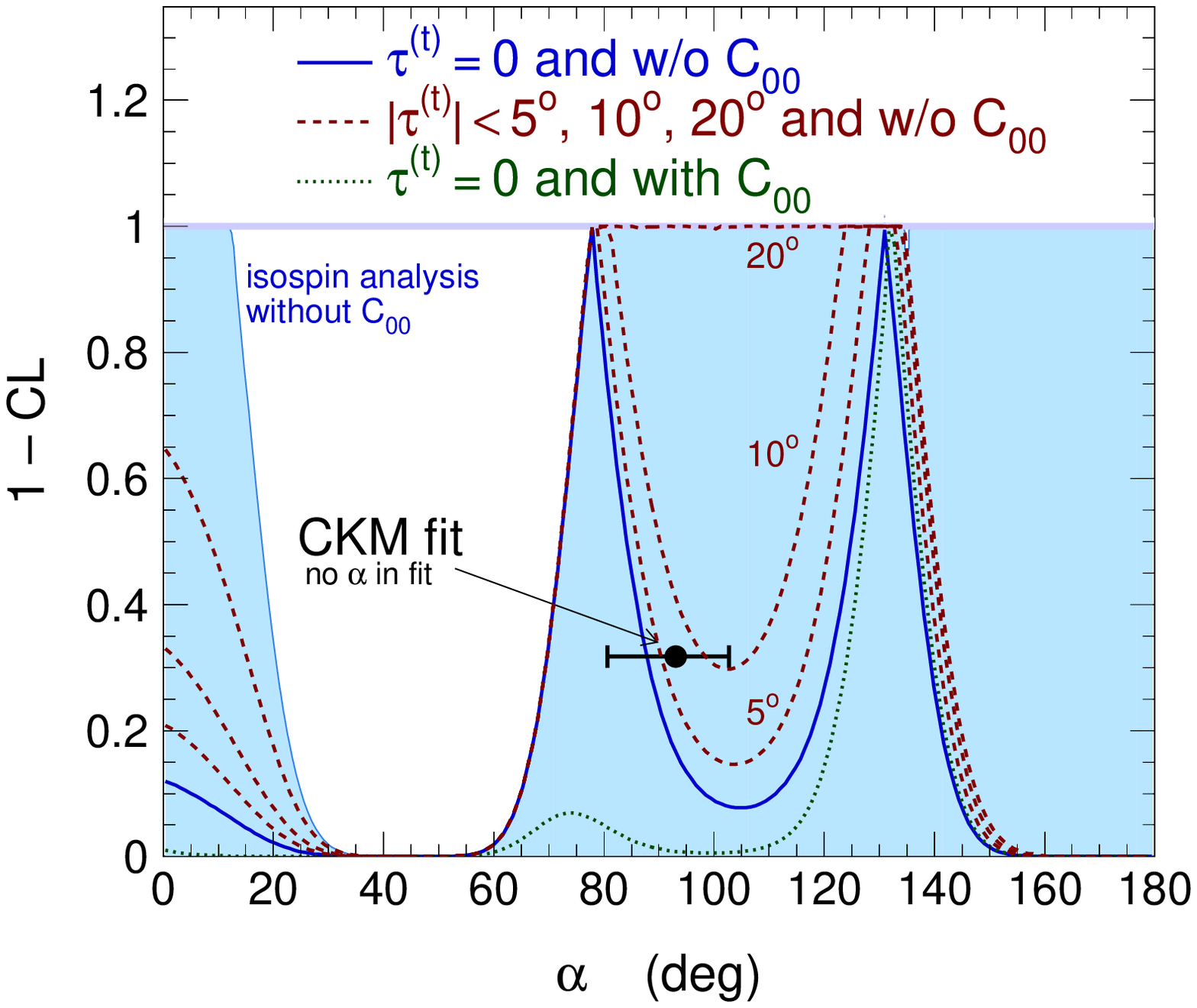}
\vspace{-0.6cm}
\caption{\underline{Left plot:} confidence level for $\alpha$ imposing $\tau = 0$
	in the t- (solid line) and c-conventions (dashed line) without using
	$C_{00}$ in the fit. The t-convention curve uses $\beta$ as an input.
        Also shown are the results of the traditional
	isospin analysis~\cite{Gronau:1990ka,Charles:2004jd} with (light shaded
	region) and without (dark shaded
	region) using $C_{00}$. The dot with $1\sigma$ error bar shows the
	predicton from the global CKM fit (not including the direct measurement
	of $\alpha$)~\cite{Charles:2004jd}.
	\underline{Right plot:} confidence level for $\alpha$ imposing $\tau =
	0$  in the t-convention with (dotted line) and without (solid line)
	using the $C_{00}$ result in the fit. Also shown are the constraints in
	the t-convention imposing $|\tau| <5^\circ$, $|\tau| <10^\circ$,
	and $|\tau| <20^\circ$ (dashed lines). The shaded
	region is the same as in the left plot.}
\label{fig:alphafit}
\end{figure*}

The confidence level (CL) of $\alpha$ obtained by imposing a constraint on
$\tau$ is shown in Fig.~\ref{fig:alphafit} using the CKMfitter
package~\cite{Charles:2004jd}.  In the left plot the curves show (see the
labels) the CL of $\alpha$ imposing $\tau = 0$ in both the t- and c-conventions
without using the $C_{00}$ measurement in the fit.  For comparison, we also show
the result of the usual isospin analysis with and without using $C_{00}$.  The
plot on the right-hand side shows the CL of $\alpha$ imposing $\tau = 0$ in the
t-convention with and without using $C_{00}$, and the constraint in the
t-convention imposing $|\tau| < 5^\circ$, $10^\circ$, and $20^\circ$.  The
restriction on $\alpha$ from a constraint $|\tau| < \tau_0$ becomes quite weak
as $\tau_0$ increases in the range $10^\circ < \tau_0 < 20^\circ$.  We can
compare our results with those of~\cite{brs}, which use as theory input an
upper bound on $\epsilon = |{\rm Im}(C^{(t)}_{\pi\pi}/T^{(t)}_{\pi\pi})|$. 
Assuming $\{ \gamma, |\arg(P_{\pi\pi}/T_{\pi\pi})|\} < 90^\circ$, we find $\sin
\tau^{(t)} < \epsilon\, \sqrt{R_{+-}}$, i.e., $\tau^{(t)} < 15.5^\circ \
(7.8^\circ)$ for the bounds considered in \cite{brs}, $\epsilon < 0.4 \ (0.2)$.

Imposing $\tau=0$ gives only two solutions with $\chi^2=0$ with the current
data, around $\alpha \sim 78^\circ$ and $132^\circ$.  The first one, which is
consistent with the Standard Model (SM) CKM fit, is disfavored by the
measurement of $C_{00}$.  While the two solutions have comparable errors for
$\tau = 0$, allowing a finite range of $\tau$ to account for subleading effects
increases the error of the $\alpha \sim 132^\circ$ solution more rapidly. 
Imposing a bound on $|\Im(C/T)|$~\cite{brs} allows, in addition to $\tau$ being
near 0, that $\tau$ is near $\pi$ (mod $2\pi$); however, the theory disfavors
the latter possibility.  It is constraining $|\tau|$ modulo $2\pi$ and not $\pi$
that makes some of the CL curves not periodic with a period of $\pi$.

These results for $\alpha$ should not be taken at face value, because in the
next Section we find that extracting $\tau$ using the SM CKM fit as an input
gives significantly larger values of $|\tau|$ than considered here.  The
implications of this are discussed below.

\section{The penguin hierarchy problem}

If the penguin amplitudes were small then the statements in (\ref{cute})  would
all hold to a good precision, and $\alpha$ could be extracted simply from
$S_{+-}$. This is known not to be the case, so the question is to  determine
which penguins are large or small.  This is complicated by the fact that, as
explained in Sec.~II, the amplitudes $T$, $C$, $P_{uc}$, and $P_{ut}$  are not
separately observable from the $B\to \pi\pi$ data alone.  They can be
disentangled using $SU(3)$ flavor symmetry and data on $B\to K\pi$, $K\Kb$,
etc.

In this section we propose to use the theory expectation for $\phi_T$ in
Eq.~(\ref{phiTC}) to test the magnitude of the penguins.  (Another test of
corrections to factorization in $B\to\pi\pi$ was proposed
in~\cite{Feldmann:2004mg}.)  We assume $\phi_T=0$, although we may learn from
other data that power corrections to tree amplitudes are sizable.  For example,
a power suppressed strong phase around $30^\circ$ is observed in $B\to D\pi$
decays~\cite{Mantry:2003uz}.

In the t-convention $P_{ut}$ (recall, $P_{ij} \equiv P_i-P_j$) contributes to
the TT in Eq.~(\ref{TT}), while in the c-convention it is $P_{uc}$.  (We choose,
for convenience, the pure tree amplitude $T_{-0}$ to be real.)  Thus, comparing
the TT in the two conventions teaches us about the relative size of $P_{ut}$ and
$P_{uc}$.  (The same information can in principle be obtained from the fit in
any one convention; this comparison makes the results more transparent.)  We use
the SM global fit to the CKM matrix that determines the weak phase
$\gamma =(59.0^{+6.4}_{-4.9})^\circ$~\cite{Charles:2004jd}. This allows
the construction
of the tree triangles in both conventions, as explained in Sec.~II.  Comparing
how flat they are, {\em i.e.}, how small the angle $\tau$ of the TT is, the
following outcomes are possible:

\begin{figure*}[t]
\includegraphics[width=\columnwidth]{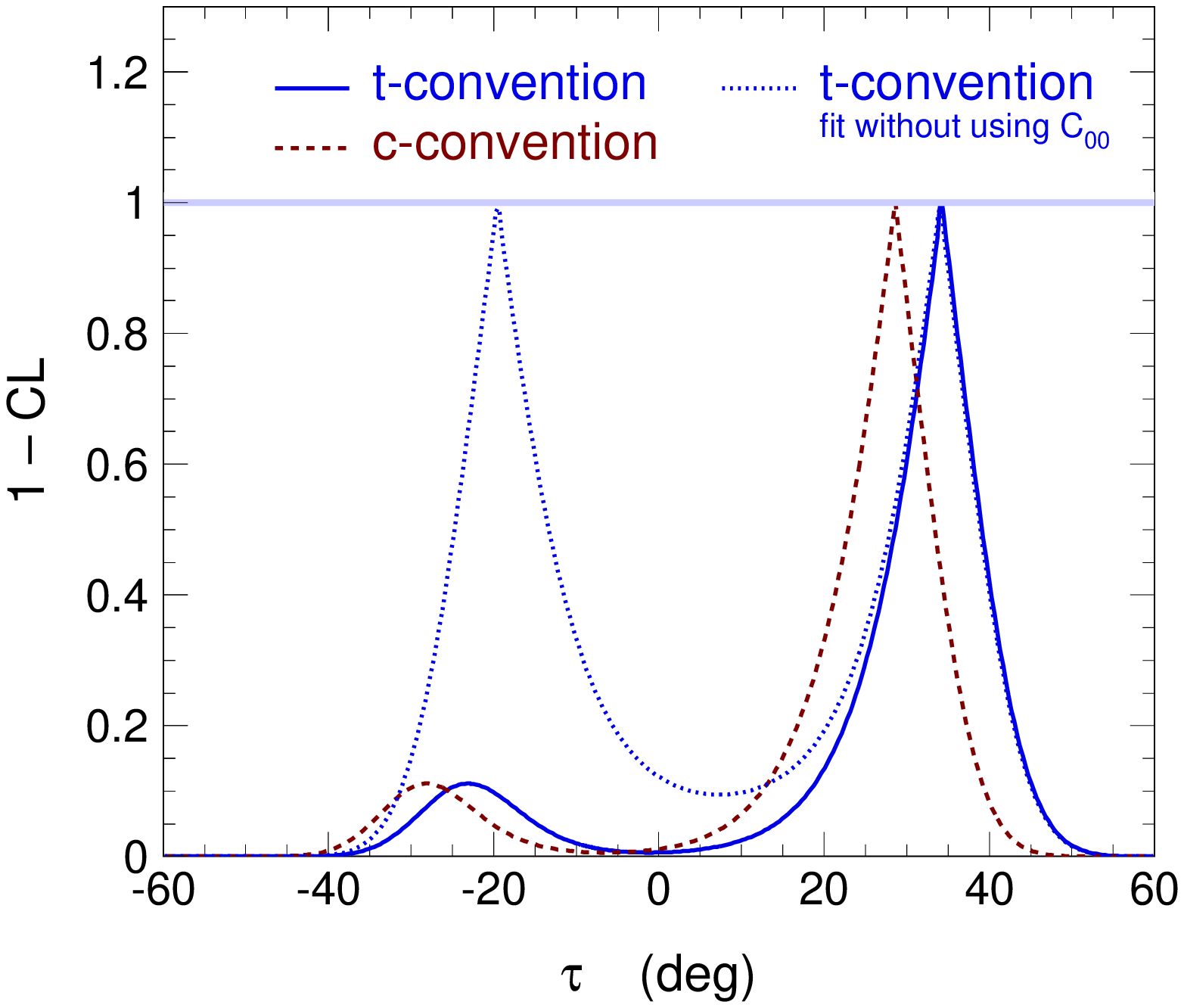} \hfill
\includegraphics[width=\columnwidth]{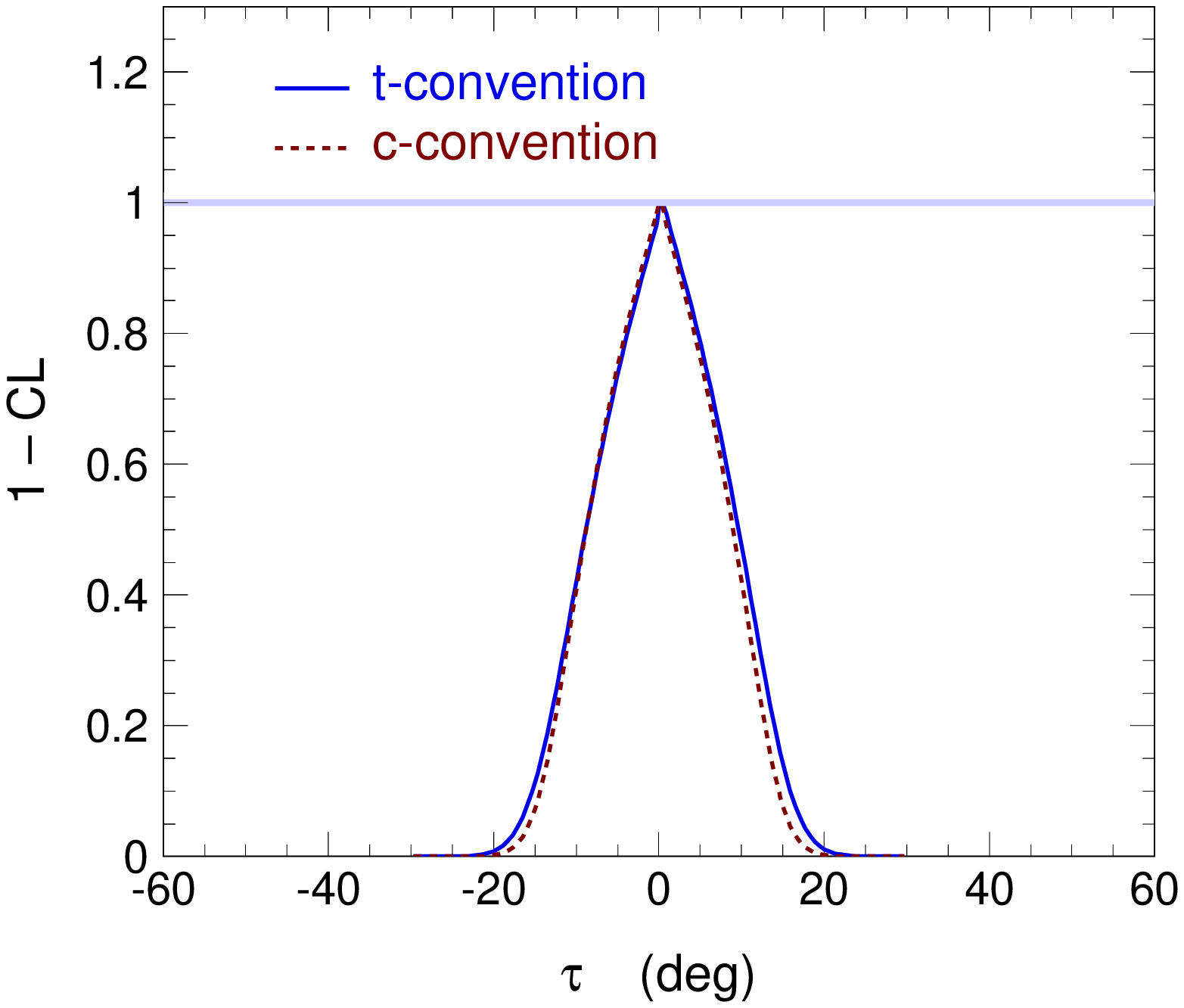}
\vspace{-0.6cm}
\caption{Confidence level plots for $\tau = \arg (T_{\pi\pi}/T_{-0})$ in
	the t- and c-conventions in $\Bb\to \pi\pi$ (left), and for
	$\Bb\to \rho\rho$ (right).}
\label{fig:tau_pirho}
\end{figure*}

\begin{itemize}

\item[(i)] $|\tau^{(t)}| \ll |\tau^{(c)}|$.  This would imply  $\Im(P_{ut}) \ll
\Im(P_{uc})$, and the likely explanation would be $|P_c| \gg |P_u| \sim  |P_t|$.

\item[(ii)] $|\tau^{(t)}| \gg |\tau^{(c)}|$.  This would imply  $\Im(P_{ut}) \gg
\Im(P_{uc})$, and the likely explanation would be $|P_t| \gg |P_u| \sim  |P_c|$.

\item[(iii)] $|\tau^{(t)}| \sim |\tau^{(c)}| \ll 1$.  This would imply  that
both $\Im(P_{ut}/T^{(0)})$ and $\Im(P_{uc}/T^{(0)})$ are small.  In this case
the likely explanation would be that $P_q/T^{(0)}$ is small for each of the
penguin amplitudes.

\item[(iv)] $|\tau^{(t)}| \sim |\tau^{(c)}| = {\cal O}(1)$ and  $|\tau^{(t)} -
\tau^{(c)}| \ll 1$.  This would imply that $\Im(P_{ut}/T^{(0)})$ and
$\Im(P_{uc}/T^{(0)})$ are both much larger than $\Im(P_{ct}/T^{(0)})$.  There
appears to be no single plausible explanation for such a case.  It may indicate
that $P_u$ (that includes weak annihilation) is large, while $P_c$ and $P_t$ are
small or have small phases.  Another, fine tuned, possibility is that both $P_c$
and $P_t$ have large but nearly equal phases.  Last, it might be that $\phi_T =
{\cal O}(1)$, indicating large corrections to the heavy quark limit.

\item[(v)] $|\tau^{(t)}| \sim |\tau^{(c)}| = {\cal O}(1)$ and $|\tau^{(t)}  -
\tau^{(c)}| = {\cal O}(1)$.  This would imply that $\Im(P_{ut}/T^{(0)})$,
$\Im(P_{uc}/T^{(0)})$, and $\Im(P_{ct}/T^{(0)})$ are all large.  In this case
the likely explanation would be that all penguins are large and comparable to
$T^{(0)}$.

\end{itemize}

Note that the $\tau^{(t)} - \tau^{(c)}$ difference is related to the
penguin-to-tree ratio,
\beq\label{deltau}
\tau^{(t)} - \tau^{(c)} = - \arg \bigg( 1 -
   \frac{|\lambda_u|}{|\lambda_c|}\,
   \frac{P_{\pi\pi}^{(t)}}{T_{\pi\pi}^{(t)}} \bigg),
\eeq
and can be determined with better precision than $\tau^{(t,c)}$ separately.

\subsection{\boldmath $B\to\pi\pi$}

Using the experimental data we can determine $\tau$ in the t- and c-conventions.
The results for the confidence levels of $\tau^{(t,c)}$ are shown in the left
plot in Fig.~\ref{fig:tau_pirho}.  At the one sigma level only one solution is
allowed (because $C_{00}$ disfavors one of the solutions at a near $2\sigma$
level).  Including $C_{00}$ in the fit drives $|\tau|$ to larger values
\beq \label{tau-val}
\tau=
\cases{\big(36^{+6}_{-8}\big)^\circ , & t-convention, \cr
  \big(30^{+6}_{-8}\big)^\circ , & c-convention. \cr}
\eeq
Note that the central values indicate rather large values for $\tau$
in both conventions.  Their difference is more accurately determined
by Eq.~(\ref{deltau}), where the fit gives
\beq\label{delta-tau-val}
\tau^{(t)} - \tau^{(c)} = \big(5.7^{+2.0}_{-1.7} \big)^\circ .
\eeq
Eqs.~(\ref{tau-val}) and (\ref{delta-tau-val}) favor scenario (iv).  While this
may have several reasons as explained above, the least fine-tuned one, {\em
i.e.},  a large $P_u$ (including weak annihilation) and smaller $P_{c,t}$
penguins (or that the $\phi_T \ll 1$ prediction receives large corrections),
would be puzzling for any approach to factorization.  At present, this is not a
very firm conclusion yet.  (Note that a similar enhancement of the $u$-penguin
amplitude is observed in $\Bb\to K\pi$ and $b\to(s\sbar)s$ decays, if the
apparent  anomalies therein are interpreted within the SM.)

\subsection{\boldmath $B\to\rho\rho$}

Since $B\to\rho\rho$ decays are dominantly longitudinally polarized, the
determination of $\alpha$ from this mode is very similar to that from
$B\to\pi\pi$, except that at the few percent level an $I=1$ amplitude may be
present~\cite{Falk:2003uq}. Using dynamical input to reduce the uncertainty of
$\alpha$ from $B\to\rho\rho$ has received little attention so far, because the
isospin bound puts tight constraints on $\alpha - \alpha_{\rm eff}$.  However,
this bound may become worse in the future, since the strong present bound is a
consequence of the fact that the isospin triangles do not close with the central
values of the current world averages.  This is a consequence of both the
branching ratios, whose central values in units of $10^{-3}$ are $\sqrt{{\cal
B}(B\to \rho^+\rho^0)} = 5.14$,  $\sqrt{{\cal B}(B\to \rho^+\rho^-)/2} = 3.87$,
and $\sqrt{{\cal B}(B\to \rho^0\rho^0)} < 1.05$ (90\%~CL), and the smallness of
$C_{\rho^+\rho^-} = -0.03 \pm 0.20$~\cite{Group:2004cx,rhorho}.  Therefore,
although at present imposing $|\tau| < 10^\circ$ does not improve the constraint
on $\alpha - \alpha_{\rm eff}$ in this mode, such a dynamical input may become
useful in the future.

In this case, the $\tau$ values in the two conventions differ by less than a
degree as shown in the right plot in Fig.~\ref{fig:tau_pirho}, giving $\tau =
(0\pm12)^\circ$.  This may tend towards the above scenario (iii).  If in the
future the measured value of the $B\to\rho^+\rho^0$ branching ratio decreases
(or that of $\rho^0\rho^0$ increases) then the pure isospin bound will become
worse, and the fit results for $\tau$ will also change.  If that fit still
favors $|\tau^{(t)}| \ll |\tau^{(c)}|$ or $|\tau^{(t)}| \sim |\tau^{(c)}| \ll 1$
[cases (i) or (iii)] then we would feel comfortable imposing a constraint on the
magnitude of $\tau^{(t)}$ to improve the determination of the CKM
angle~$\alpha$.

\section{Conclusions}

The tree amplitudes in $B\to \pi\pi$ decays can be computed in an expansion of
$\lqcd/m_b$ using factorization.  In the heavy quark limit the strong phase
between the tree amplitudes is suppressed, which may help to improve the
determination of the weak phase $\alpha$.
Using this theory input as an additional constraint in the fit for $\alpha$,
requires some understanding of the power corrections and penguin amplitudes.

While the present measurement of $C_{00}$ does not provide a significant
determination of $\alpha$ from the $\B\to\pi\pi$ isospin analysis, it provides
useful information about the hadronic amplitudes.  The determination of $\alpha$
using the central values of the present data with $C_{00}$ replaced by the
assumption of a flat TT gives a solution near the isospin bound.  While a
$|\tau^{(t)}| < 5^\circ$ or $10^\circ$ theoretical bound is quite powerful to
constrain $\alpha$, allowing for larger deviations from the heavy quark limit
$(|\tau^{(t)}| < 20^\circ)$ reduces significantly the predictive power of the
constraint on $\alpha$.  The present $C_{00}$ result, however, disfavors being
at the isospin bound at about the $2\sigma$ level.  This observation is
exhibited by the like-sign $C_{+-}$ and $C_{00}$ measurements, whereas the
opposite signs of the $P_{\pi\pi}$ terms in the $\pi^+\pi^-$ and $\pi^0\pi^0$
amplitudes would imply opposite signs for $C_{+-}$ and $C_{00}$ if the tree
triangle was flat.

We proposed a comparison of fits that can give information about the relative size
of the penguins, using only $\pi\pi$ data and the global fit for $\gamma$.  While the
present data is not yet precise enough to give firm conclusions, its most likely
implication is that not only the charm (nor the top) penguins in  $B \to \pi\pi$
are large, but so are the up penguins (including terms proportional to $V_{ub}$
that are power suppressed in the heavy quark limit), thus one may not be
able to use theory instead of $C_{00}$.  On the other hand, for $B\to
\rho\rho$  decay, it may well be the case that the data will continue to favor
$|\tau^{(t)}| \sim |\tau^{(c)}| \ll 1$ or $|\tau^{(t)}| \ll |\tau^{(c)}|$, in
which case the theory can be useful to reduce the error on $\alpha$ without a
measurement of $C_{00}$.

\acknowledgments

We thank Christian Bauer, Andy Cohen, Marco Ciuchini, and Iain Stewart for
useful discussions.
Special thanks to J\'er\^ome Charles for helpful comments and for pointing out a
mistake in our earlier numerical results.
We thank the Institute for Nuclear Theory at the University
of Washington for its hospitality and partial support while some of this work
was completed. Z.L.\ thanks the particle theory group at Boston University for
its hospitality while part of this work was completed.
This work was supported in part by the Director, Office of Science,
Office of High Energy and Nuclear Physics, Division of High Energy
Physics, of the U.S.\ Department of Energy under Contract
DE-AC02-05CH11231 and by a DOE Outstanding Junior Investigator award
(Z.L.); and by the U.S.\ Department of Energy under cooperative research
agreement DOE-FC02-94ER40818 (D.P.).

\end{document}